\documentclass[10pt]{article}
\usepackage{graphicx,amssymb, amstext, amsmath, epstopdf, booktabs, verbatim, gensymb, geometry, appendix, lmodern,bchart,pgfplots,url,hyperref}
\usepackage{chngcntr}
\usepackage{color, colortbl}
\usepackage{appendix}
\usepackage{wrapfig}
\usepackage{lscape}
\usepackage{rotating}
\usepackage{epstopdf}

\hypersetup{
    colorlinks=true,
    linkcolor=blue,
    filecolor=blue,      
    urlcolor=blue,
    bookmarks=true,
    citecolor=blue,
    pdfpagemode=FullScreen,
    }

\counterwithout{figure}{section}
\counterwithout{table}{section}

\pgfplotsset{width=7cm,compat=1.8}

\newcommand*\Title{KOALA Project Report 3}
\newcommand*\cpiType{--- KOALA Project Report 3}
\newcommand*\Date{January 2019}

\title{Are Children Fully Aware of Online Privacy Risks and How Can We Improve Their Coping Ability?}

\author{Ge Wang \\ Jun Zhao \\ Nigel Shadbolt}
\date{\today}

\usepackage{cpi} 

\begin{document}

\begin{titlepage}
\maketitle
\end{titlepage}

\linespread{1.15} 

\begin{executive}

The age of children adopting digital technologies, such as tablets or smartphones, is increasingly young. However, children under 11 are often regarded as too young to comprehend the concept of online privacy. Limited research studies have focused on children of this age group. In the summer of 2018 we conducted 12 focus group studies with 29 children aged 6-10 from Oxfordshire primary schools. Our research has shown that children have a good understanding of certain privacy risks, such as \textit{information oversharing} or \textit{avoiding revealing real identities online}. They could use a range of descriptions to articulate the risks and describe their risk coping strategies. However, at the same time, we identified that children had less awareness concerning other risks, such as \textit{online tracking} or \textit{game promotions}.

Inspired by Vygotsky's Zone of Proximal Development (ZPD), this study has identified critical knowledge gaps in children's understanding of online privacy, and several directions for future education and technology development. We call for attention to the needs of raising children's awareness and understanding of risks related to online recommendations and data tracking, which are becoming ever more prevalent in the games and content children encounter. We also call for attention to children's use of language to describe risks, which may be appropriate but not necessarily indicate a full understanding of the threats. 

\frame{
\textbf{Four key findings}
\begin{enumerate}
    \item Children under 11 \textbf{cared about their privacy online}, and were sensitive to, who might access their sensitive information (e.g. real names, age, location etc), and applied a range of techniques to safeguard this space.
    
    \item However, children still \textbf{need help to fully understand online privacy risks}, especially those associated with implicit personal data collection and use. Children were more likely to associate new videos promoted to them on YouTube as ``auto-plays'' and struggled to recognise ``recommendations'' and links to their personal data privacy.
    
    \item Children may \textbf{use the correct language to describe risks, without fully comprehending the risks}: this may lead to effective risk coping because children associated the risks with perceived threats; however, children may also try to make sense out of the context using their knowledge or experiences, and not take an effective action. 
    
    \item When children struggled to fully understand the implications of certain risks, \textbf{they often relied on their own or their friends' experiences to make risk-related decisions}. They would opt for ``play and see'', which is consistent to their nature of enjoying explorations; however, they could also leave themselves exposed to inappropriate content or online baiting. This indicates the need to have deeper discussions about the implications of losing control of our data online and reinforce learning by the sharing of experiences.

\end{enumerate}
}

\end{executive}

\section{Introduction}
Today, children are spending more time online than with other media sources, such as televisions or offline video games~\cite{ofcom2017,livingstone2017children}. Among the many kinds of devices now connected to the Internet, mobile devices (such as tablet computers or smartphones) have become the primary means by which children go online~\cite{livingstone2017children}. In the UK, 35\% of children aged five to seven and 52\% of children aged eight to eleven have been reported to have their own tablets  (see Figure 1)~\cite{ofcom2017}; while in the US, ownership of tablets by children in this age group grew five-fold between 2011 and 2013~\cite{commonsense2013}. Children under five are also using smartphones and tablets more often, as the category of apps designed for younger kids continues to expand rapidly~\cite{ofcom2017, lifewire}. 

\begin{figure}[h!]
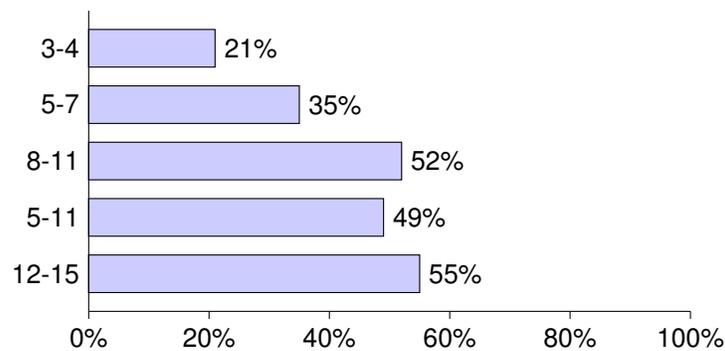

\centering
\begin{bchart}[step=20,max=100,unit=\%]
        \bcbar[label=3-4]{21}
            \smallskip
        \bcbar[label=5-7]{35}
            \smallskip
        \bcbar[label=8-11]{52}
            \smallskip
        \bcbar[label=5-11]{49}
            \smallskip
        \bcbar[label=12-15]{55}
        
\end{bchart}
\caption{Percentage of Children Owning Their Own Tablets In Proportion to the Total Number of Tablet Accessible to the Household by age~\cite{ofcom2017}}
\end{figure}\label{fig:devices}

However, for children aged between 6 and 10, their awareness of online privacy has been seldom studied in existing literature. It's probably due to the expectation that children of this age cannot comprehend the concept of privacy, and therefore their online safety and privacy should be safeguarded by the trusted grown-ups. However, our previous reports~\cite{DBLP:journals/corr/abs-1809-10841,DBLP:journals/corr/abs-1809-10944} have shown that although parents largely take a protective approach to restrict and monitor what their children can or cannot access on their children's devices, parents do not always fully comprehend all risks. Furthermore, a more recent study led by the LSE has shown that parents struggle to seek for help to mediate their children's use of digital devices, and they largely rely on themselves to figure things out~\cite{livingstone2018parentsb}. 

Therefore, the goal of the KOALA project is to, through a series of studies directly interacting with children and their families, to achieve a deeper understanding of the following questions:
\begin{itemize}
    \item How well can children aged 6-10 understand online privacy risks related to their use of tablet computers?
    \item What are the barriers for their parents to mediate children's use of technologies and recognition of risks?
\end{itemize}

Between June and August 2018 we conducted 12 focus group interviews with 29 children, aged 6-10, from UK schools. We found that children in our study had a good understanding of risks related to \textit{inappropriate content, the approach of strangers}, and \textit{oversharing of personal information online}. However, they struggled to fully understand and describe risks related to \textit{online game/video promotions and personal data tracking}. Moreover, children's risk coping strategies depended on their understanding of the risks and their previous experiences: effective risk strategies were applied only if children recognised certain risks or when they \textit{felt} something untoward. These findings demonstrate the importance of learning about potential risks through a multitude of channels, such as school, parents, friends, siblings, and personal experiences~\cite{zhang2017cyberheroes,hashish2014involving}.
\section{Methodology}
We chose the focus group method to elicit children's responses to a collection of hypothetical scenarios that reflect different types of explicit and implicit threats to children's online personal data privacy. 

Each focus group study contained four parts, including a warm-up and introduction session, a sharing of favourite apps, a walk-through of three hypothetical risk scenarios, and finally an open-ended session about issues not so far discussed. 

Children were encouraged to share their personal experience related to the scenarios, by responding to questions like \textit{whether this has happened to them} and \textit{what they did}. This enabled us to listen to children's descriptions of their experiences not covered in the scenarios. 

The hypothetical scenarios used in the study have been created into story cards that can be reused and shared in the classrooms or at home. They can be found in the Appendix towards the end of the report, accompanied by descriptions about the privacy risks related to each card. Parents and teachers can also download our story cards from \href{https://sites.google.com/view/koala-project-ox/toolkits}{our project web site} and use these cards to inspire conversations with children about online privacy issues.

\section{Key Findings}

Through careful data analysis of 20-hour interview recordings, we have achieved some positive findings with respect to children's level of awareness. However, at the same time, we have also identified critical gaps in children's online privacy knowledge. Children sometimes may be able to capture certain risks with the right language, however, they may not have fully comprehended the risks. These are critical insights that should draw further attention to families and schools when applying risk mediation to children. The terms used by the participant children and their abilities to apply these terms to different types of risk context has provided valuable inputs to our future design of technologies to facilitate children's risk coping abilities. 

\subsection{Children care about online privacy risks}
Our results reinforced that, like teenagers and younger children, our participants valued the positive experiences of being online and keeping their personal space online~\cite{lwin2008protecting}. They enjoyed going online so that they could keep in touch with friends, learn new things or just have fun.

At the same time, children felt `annoyed', `surprised' or `angry' when they felt coerced or not in control. As a result, children cared about, and were sensitive to, who might access their sensitive information (e.g. real names, age, location etc), and applied a range of techniques to safeguard this space, such as by verifying identities through face-to-face interactions or avoiding using real names as usernames. 

\subsection{Children still need help with certain privacy risks} 

Children demonstrated a strong consciousness of their online identity and the importance of avoiding sharing their real identity or over-sharing their personal information. In these cases, children applied various effective strategies to protect their sensitive personal information, such as `making up a new name', and asking their friend to verify the online invitation in person. However, in contrast to explicit privacy risks, children struggled to associated ``online promotions'' with losing control of personal information. 

Several children correctly discussed their interpretation of how YouTubers may try to `persuade' them to watch their videos in order to gain `money' or `more subscribers'. However, none of these children expressed resistance to these video promotions. They treated it as if this is how the Internet works. 

Only a few children (3/12 focus groups) recognised how new videos might be related to personalised recommendations. However, they struggled to understand who was performing these recommendations, and as a result, they were less sure about the consequences of their privacy. As a result, the children usually applied the play-and-see strategy to assess the content or apps they came across, without any more complex reasoning.

\subsection{Children's ability to describe risks depends on their actual understanding of risks} 

Our study is partially inspired by Vygotsky's Zone of Proximal Development (ZPD), and therefore, we really focused on understanding children's current knowledge by observing the language they used to describe risks. We observed that some terms were repeatedly used by children across different situations --- children were able to describe risks accurately when they could recognise the actual risks. However, when they had only a vague understanding of the risks, they struggled to describe things consistently or to provide a good explanation of what they meant.

When children only made a partial sense of certain types of risks, their use of terms can be inconsistent. For example, although some children were able to describe the scenario of new videos being presented to Bertie using words like `people trying to make money' or `them trying to make you watch more', they struggled to explain who these `people' are and how information might be transmitted to these `[YouTube] channel people' in this context. 

Another example is the term `hacking', which has been used by children across different focus groups, but in fact with very different meanings. For example, when trying to explain why a new video was shown following a previous video, children used `hacking' to mean `someone stole my data', `take your account', or `steal from house'. `Hacking' has been used by the same group of children to make sense of other scenarios, including in-app pop-ups or data tracking.

\subsection{Children struggle to comprehend the implications of risks}

Children could miss risks due to a lack of knowledge, or due to their past experiences, which did not lead to any direct consequences related to the risks. For those children who interacted with certain technologies and experienced no implications before, they would tend to be more (over-) confident with technologies.

Children from 7 focus groups treated online video promotions as part of an `autoplay' function of the platform, without questioning how the new content might be presented to them. 12/29 children demonstrated trust in the content provided by their familiar YouTubers. As a result, children reported having been exposed to unexpected content and online baiting. These same children reported that they often saw upsetting content online (e.g. `sometimes in autoplay, it comes up with these really freaky ones like pictures of dead people').

When children not fully comprehending the risks, we also observed that a child's or their peers' personal experiences had a strong influence upon their decision making, even though they didn't always understand what may pose threats to them.

\newpage

\frame{
\textbf{Take-Home Messages For Parents \& Caretakers}
\begin{enumerate}
    \item Continue to talk to children about being careful online, because they are indeed facing challenges in their use of digital devices every day and they really care about protecting their personal space online!
    
    \item Pay attention to how children describe risks or things that made them uncomfortable online. This is a good start to get to know what they actually mean, and it's always good to follow this up with a question like ``what do you mean?''.
    
    \item Do not be afraid of talking about digital technologies with children. They would enjoy learning with you and sharing their knowledge with you.
    
    \item If you are indeed struggling, please take a look at the resources we recommend below. It's never too late to get to know what your child knows about online privacy and how you may help them or how you may help each other!
\end{enumerate}
}

\section{Resources}
\begin{itemize}
    \item \href{commonsensemedia.org}{commonsensemedia.org} would be a good place for finding out additional information about which mobile app would be good for your child, such as Minecraft, Roblox, etc. 
    
    \item \href{https://www.net-aware.org.uk}{NSPCC Net Aware} provides good evidence-based advice on apps that are most popular among children.
    
    \item \href{childnet.com}{childnet.com} has good resources for talking about online privacy with younger children, such as the story of \href{https://www.childnet.com/resources/smartie-the-penguin}{Smartie the Penguin} or the story book of  \href{https://www.childnet.com/resources/digiduck-stories}{Digiduck}.
    
    \item Finally, our KOALA Hero is always here to help, either through the \href{https://sites.google.com/view/koala-project-ox/toolkits/cards}{story cards} or regularly updated \href{https://sites.google.com/view/koala-project-ox/reports-and-publications}{public reports}. 
\end{itemize}

\section{Acknowledgement} 

We thank all the schools, families and children who have contributed their time and knowledge to our study. Without their support, this study would not have been successful!

The report is based on a full academic publication \href{https://arxiv.org/abs/1901.10245}{`I make up a silly name': Understanding Children's Perception of Privacy Risks Online}, as part of the project of KOALA (https://sites.google.com/view/koala-project-ox/): Kids Online Anonymity \& Lifelong Autonomy, funded by EPSRC Impact Acceleration Account Award, under the grant number of EP/R511742/1.

\section{Contact} 
\begin{itemize}
    \item Jun Zhao: jun.zhao@cs.ox.ac.uk
    \item Nigel Shadbolt: nigel.shadbolt@cs.ox.ac.uk
\end{itemize}

\appendix
\section{Three Hypothetical Risk Scenarios}
Our scenarios were carefully designed to contrast explicit versus implicit data collection in a familiar versus unfamiliar technology context. Previous research has shown that children under 11 particularly struggle to understand risks posed by technologies or comprehend the context of being online~\cite{zhang2016nosy,kumar2018cscw}. They have explored how children responded to \textit{explicit} data requests such as in-app pop-ups; while our study also looks into children's awareness and perception of \textit{implicit} data access through third-party tracking, which leads to personalised online promotions.

Story 1 --- Implicit video promotions are widely found in applications like social video sharing platforms, which can be based on personal viewing history or a viewers' interests. Online promotions on such platforms are a significant means by which children discover games or video channels, material that is not always appropriate for their age or developmental needs~\cite{livingstone2017children}. Therefore, story 1 aimed to assess how much children are aware of the video promotion behaviours of online platforms, some of which could be based on children's online activities, including the videos they have watched or the games they enjoy playing.

Story 2 --- In-app pop-ups can explicitly prompt children for personal information (such as names, age or voices) before they can continue with the game. Story 2 aimed to assess children's awareness of \textit{explicit} stranger danger and in-app game promotions, which can be personalised based on their online data.

Story 3 --- The large number of applications (`apps') that can be downloaded for free are a major way by which children interact with these devices. Currently these `free' apps are largely supported by monetisation of user's personal information~\cite{acquisti2016economics,kummer2016private}. A large amount of personal information and online behaviour may be collected from children's apps and shared with third party online marketing and advertising entities~\cite{reyes2017our}. This scenario is designed to examine how children perceive and feel about these risks. 

\begin{sidewaysfigure}[ht]
    \includegraphics[width=\textwidth]{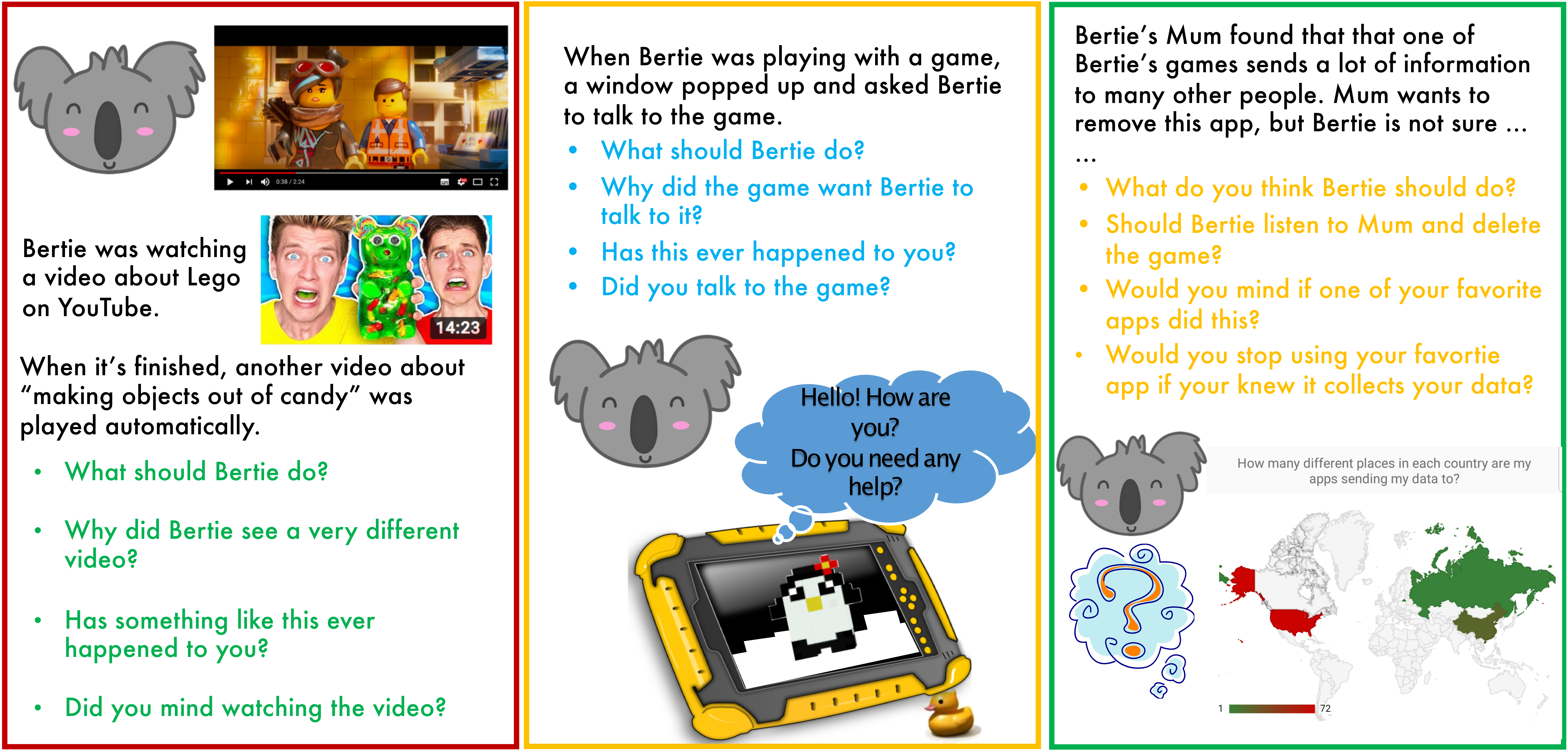}
\end{sidewaysfigure}

\bibliographystyle{SIGCHI-Reference-Format}

\begin{thebibliography}{00}


\ifx \showCODEN    \undefined \def \showCODEN     #1{\unskip}     \fi
\ifx \showDOI      \undefined \def \showDOI       #1{{\tt DOI:}\penalty0{#1}\ }
  \fi
\ifx \showISBNx    \undefined \def \showISBNx     #1{\unskip}     \fi
\ifx \showISBNxiii \undefined \def \showISBNxiii  #1{\unskip}     \fi
\ifx \showISSN     \undefined \def \showISSN      #1{\unskip}     \fi
\ifx \showLCCN     \undefined \def \showLCCN      #1{\unskip}     \fi
\ifx \shownote     \undefined \def \shownote      #1{#1}          \fi
\ifx \showarticletitle \undefined \def \showarticletitle #1{#1}   \fi
\ifx \showURL      \undefined \def \showURL       #1{#1}          \fi

\bibitem{commonsense2013}
 2013.
\newblock {\em Zero to Eight: Children's Media Use in America 2013}.
\newblock {T}echnical {R}eport. Common Sense Media.
\newblock
\showURL{%
\url{https://www.commonsensemedia.org/research/zero-to-eight-childrens-media-use-in-america-2013}}


\bibitem{lifewire}
 2019.
\newblock Must Have Apps for Kids Under Five.
\newblock
  \url{https://www.lifewire.com/great-apps-for-kids-5-and-under-4152990}.
  (2019).
\newblock
\newblock
\shownote{Accessed: 2018-12-15.}


\bibitem{acquisti2016economics}
{Alessandro Acquisti}, {Curtis~R Taylor}, {and} {Liad Wagman}. 2016.
\newblock \showarticletitle{The economics of privacy}.
\newblock {\em Journal of Economic Literature\/} {52}, 2 (2016).
\newblock


\bibitem{hashish2014involving}
{Yasmeen Hashish}, {Andrea Bunt}, {and} {James~E Young}. 2014.
\newblock \showarticletitle{Involving children in content control: a
  collaborative and education-oriented content filtering approach}. In {\em
  Proceedings of the 32nd annual ACM conference on Human factors in computing
  systems}. ACM, 1797--1806.
\newblock


\bibitem{kumar2018cscw}
{Priya Kumar}, {Shalmali~Milind naik}, {Utkarsha~Ramesh Devkar}, {Marshini
  Chetty}, {Tamara~L. Clegg}, {and} {Jessica Vitak}. 2018.
\newblock \showarticletitle{No Telling Passcodes Out Because They're Private?:
  Understanding Children's Mental Models of Privacy and Security Online}. In
  {\em Proceedings of ACM Human-Computer Interaction (CSCW '18 Online First)}.
  ACM.
\newblock


\bibitem{kummer2016private}
{Michael~E Kummer} {and} {Patrick Schulte}. 2016.
\newblock \showarticletitle{When private information settles the bill: Money
  and privacy in Google's market for smartphone applications}.
\newblock {\em ZEW-Centre for European Economic Research Discussion Paper\/}
  16-031 (2016).
\newblock


\bibitem{livingstone2017children}
{Sonia Livingstone}, {Julia Davidson}, {Joanne Bryce}, {Saqba Batool}, {Ciaran
  Haughton}, {and} {Anulekha Nandi}. 2017.
\newblock {\em Children's online activities, risks and safety: a literature
  review by the UKCCIS evidence group}.
\newblock {T}echnical {R}eport. UKCCIS evidence group.
\newblock


\bibitem{livingstone2018parentsb}
{Sonia Livingstone} {and} {Kjartan Olafsson}. 2018.
\newblock {\em When do parents think their child is ready to use the internet
  independently?}
\newblock {T}echnical {R}eport. LSE.
\newblock


\bibitem{lwin2008protecting}
{May~O Lwin}, {Andrea~JS Stanaland}, {and} {Anthony~D Miyazaki}. 2008.
\newblock \showarticletitle{Protecting children's privacy online: How parental
  mediation strategies affect website safeguard effectiveness}.
\newblock {\em Journal of Retailing\/} {84}, 2 (2008), 205--217.
\newblock


\bibitem{ofcom2017}
{ofcom.org}. 2017.
\newblock Children and parents: media use and attitude report.
\newblock   (2017).
\newblock
\showURL{%
\url{https://www.ofcom.org.uk/__data/assets/pdf_file/0020/108182/children-parents-media-use-attitudes-2017.pdf}}


\bibitem{reyes2017our}
{Irwin Reyes}, {Primal Wiesekera}, {Abbas Razaghpanah}, {Joel Reardon}, {Narseo
  Vallina-Rodriguez}, {Serge Egelman}, {and} {Christian Kreibich}. 2017.
\newblock \showarticletitle{``Is Our Children's Apps Learning?'' Automatically
  Detecting COPPA Violations}. In {\em Workshop on Technology and Consumer
  Protection (ConPro 2017), in conjunction with the 38th IEEE Symposium on
  Security and Privacy}.
\newblock


\bibitem{zhang2017cyberheroes}
{Leah Zhang-Kennedy}, {Yomna Abdelaziz}, {and} {Sonia Chiasson}. 2017.
\newblock \showarticletitle{Cyberheroes: The design and evaluation of an
  interactive ebook to educate children about online privacy}.
\newblock {\em International Journal of Child-Computer Interaction\/} (2017).
\newblock


\bibitem{zhang2016nosy}
{Leah Zhang-Kennedy}, {Christine Mekhail}, {Yomna Abdelaziz}, {and} {Sonia
  Chiasson}. 2016.
\newblock \showarticletitle{From Nosy Little Brothers to Stranger-Danger:
  Children and Parents' Perception of Mobile Threats}. In {\em Proceedings of
  the The 15th International Conference on Interaction Design and Children}.
  ACM, 388--399.
\newblock


\bibitem{DBLP:journals/corr/abs-1809-10944}
{Jun Zhao}. 2018.
\newblock \showarticletitle{Are Children Well-Supported by Their Parents
  Concerning Online Privacy Risks, and Who Supports the Parents?}
\newblock {\em CoRR\/}  {abs/1809.10944} (2018).
\newblock
\showURL{%
\url{http://arxiv.org/abs/1809.10944}}


\bibitem{DBLP:journals/corr/abs-1809-10841}
{Jun Zhao}, {Ulrik Lyngs}, {and} {Nigel Shadbolt}. 2018.
\newblock \showarticletitle{What privacy concerns do parents have about
  children's mobile apps, and how can they stay SHARP?}
\newblock {\em CoRR\/}  {abs/1809.10841} (2018).
\newblock
\showURL{%
\url{http://arxiv.org/abs/1809.10841}}


\end{thebibliography}

\end{document}